A fuzzy Multi-Criteria Decision Making approach for Exo-Planetary Habitability

**J. M. Sánchez-Lozano[1], A. Moya[2] and J. M. Rodríguez-Mozos[3]**

(1) University Centre of Defence at the Spanish Air Force Academy, MDE-UPCT. Dept. of Sciences and Computing. 30720. San Javier, Spain. E-mail: juanmi.sanchez@cud.upct.es

(2) Technical University of Madrid (UPM). School of Industrial Design and Engineering. Dept. Applied Physics. 28012. Madrid. Spain. E-mail: a.moya@upm.es

(3) University of Granada (UGR). Dept. Theoretical Physics and Cosmology. 18071. Granada. Spain. E-mail: josemaria.rodriguezmozos@gmail.com

## Abstract

Nowadays, we know thousands of exoplanets, some of them potentially habitable. Next technological facilities (JWST, for example) have exoplanet atmosphere analysis capabilities, but they also have limits in terms of how many targets can be studied. Therefore, there is a need to rank and prioritize these exoplanets with the aim of searching for biomarkers. Some criteria involved, such as the habitability potential of a dry-rock planet versus a water-rich planet, or a potentially-locked planet versus a tidally-locked planet, are often vague and the use of the fuzzy set theory is advisable. We have applied a combination of Multi-Criteria Decision-Making methodologies with fuzzy logic, the Fuzzy Reference Ideal Method (FRIM), to this problem. We have analyzed the habitability potential of 1798 exoplanets from TEPCat database based on a set of criteria (composition, atmosphere, energy, tidal locking, type of planet and liquid water), in terms of their similarity to the only ideal alternative, The Earth. Our results, when compared with the probability index SEPHI, indicate that Kepler-442b, Kepler-062e/f, and LHS_1140b are the best exoplanets for searching for biomarkers, regardless its technical difficulty. If we take into account current technical feasibility, the best candidate is TRAPPIST-1e.

## 1. Introduction

The problem of assessing the habitability of potential habitable exoplanets was faced decades before confirming their existence. Thanks to advances in fields such as biology or modern geology, the main characteristics that should satisfy a potentially habitable exoplanet began to be analyzed in the last century (Wallace, 1903). In fact, the first definition of Habitable Zone (HZ) dates back to the middle of the 20th century (Huang, 1960). According to Gaidos et al. (2005), the HZ involves diverse factors as semi-major axis, geological activity, and water, carbon and nitrogen presence. However, that concept is in constant evolution. Barnes et al. (2015), for example, assume that an exoplanet is habitable not only if its surface is telluric, but also if it is in the HZ and its emitted flux is also constrained by certain conditions. However, others studies have suggested that an exoplanet can be positioned outside the HZ but, even so, it can be a good candidate to harbor life (Irwin & Schulze-Makuch, 2011; Heller & Armstrong, 2014).

It is with the discovery of the first exoplanets (Wolszczan & Frail, 1992; Mayor & Queloz, 1995) and the large growth-rate of the inclusion of new exoplanets to the catalogs when the necessity of disentangling between potentially habitable exoplanets or not were clear. The exoplanet-discovery explosion coming from dedicated space missions such as CoRoT (ESA- other countries, Baglin et al., 2006), Kepler (NASA, Borucki et al, 2010), and TESS (NASA, Ricker et al., 2014) made this necessity a must. We must rank and prioritize, using the information mainly available, among the thousands of known exoplanets for the next step: searching for biomarkers with the new facilities such as JWST, for example. Current and future space missions (ESA – CHEOPS, Fortier et al., 2014; ESA – Plato, Rauer et al., 2014) only reaffirm this must, since the number of discovery will increase dramatically.

In this effort, numerous habitability indexes have been designed in the last decades. At the beginning of this century, a five-scale rating system (re-called Plausibility Of Life) to cover the range of plausibilities for life on other worlds (major planetary bodies and satellites in the Solar System) was proposed (Irwin & Schulze-Makuch, 2001). That groundbreaking study, through minimal generic definitions of life, established a specific set of criteria such as the presence of a fluid medium, presence of source of energy, and presence of constituents and conditions compatible with polymeric chemistry.

A decade later, with the intention of delving into exoplanet classification techniques based on their habitability, the Earth Similarity Index (ESI) and Planetary Habitability Index (PHI) were designed (Schulze-Makuch et al., 2011). The first one (ESI) is based on four physical parameters (radius, density, escape velocity and surface temperature), while the criteria which define the second one index (PHI) are the presence of a stable substrate, appropriate chemistry, available energy, and the potential for holding a liquid solvent.

Subsequent to this work, the study proposed by Irwin et al. (2014) evaluated the possibility of biological complexity on other worlds by estimating the occurrence of complex life in our galaxy, the Milky Way. As a consequence of that, the concept of Biological Complexity Index (BCI), which is based on geophysical, energy, substrates, temperature and age parameters, was defined.

In the last five years more studies have continued developing new indexes such as the Mars Similarity Index, MSI, focused on identifying planets that might be habitable but under extreme conditions of life (Kashyap & Gudennavar, 2017; Kashyap et al., 2017), the Cobb-Douglas Habitability Score (CDHS) based on a production function and physical parameters such as radio, density, escape velocity and surface temperature (Bora et al., 2016) which has been recently recomputed (Basak et al., 2020) with the imputed eccentricity values by the 'Constant Elasticity Earth Similarity Approach (CEESA), or the Statistical-likelihood Exo-Planetary Habitability Index (SEPHI) which applies likelihood functions to estimate the habitability potential using seven inputs of both planets and their stars and condensing all this information in four sub-indexes (Rodríguez-Mozos & Moya, 2017). In this work, the selected inputs are related to the main physical information the planetary discovery projects provided. For example, they are the list of Plato Data Products this mission plan to offer the community. Therefore, the selection criteria are those with the largest impact on potential habitability that can be estimated using these inputs. The main weaknesses of the SEPHI index are the references used for defining the likelihood functions (Early Mars, The Earth, etc.).

This brief review demonstrates how, through probability or production functions, statistical analysis, etc., it is possible to evaluate or classify exoplanets in terms of their habitability or capacity to harbor life. In such cases, the starting point is always common, a list of exoplanets or celestial bodies to be analyzed, and the values of the physical parameters which have influence in the assessment of these bodies. Such a characteristic constitutes what, in the field of decision theory, is so-called the decision matrix of alternatives and criteria.

An alternative and complementary way to the robust statistical analysis of SEPHI for analyzing known exoplanets for ranking them in terms of their potential habitability is to use Multi-Criteria Decision Making (MCDM) techniques. This alternative can be faced thanks to the current large data set of well-known potentially habitable exoplanets, and it is a must since in the near future the number of new discoveries will grow and the next generation instruments won't be able to study all of them. MCDM is a process for selecting among alternative courses of action, based on a set of criteria, to achieve one or more objectives (Simon, 1960). Therefore, a MCDM problem consists in a collection of alternatives that must be evaluated based on certain criteria. Since middle of last century a large number of MCDM methodologies have emerged. Among the most prominent the following could be mentioned: PROMETHEE (Preference Ranking Organization METHod for Enrichment of Evaluations) methodology developed by Brans et al. (1984), ELECTRE (ELimination and Choice Expressing Reality) from the French school (Roy, 1968), AHP method (Analytic Hierarchy Process) and ANP method (Analytic Network Process) both created by Thomas Saaty from the American school (Saaty, 1980; Saaty, 1996), TOPSIS (Technique for Order Performance by Similarity to Ideal Solution) method developed by Hwang & Yoon (1981), VIKOR method (Opricovic, 1998) or, more recently, RIM (Reference Ideal Method) method designed by Cables et al. (2016).

Although these decision techniques have been widely used in multiple disciplines (Mardani et al., 2015), their application to the field of astronomy, astrophysics, and space sciences is fairly recent. For example, these methodologies were applied to solve complex problems in NASA space exploration projects (Tavana, 2003; 2004), and AHP and TOPSIS methodologies allowed to evaluate preference models for human exploration on Mars (Tavana & Hatami-Marbini, 2011). Such combination of methodologies (AHP-TOPSIS) was utilized to assess potentially dangerous near-Earth objects (Sánchez-Lozano & Fernández-Martínez, 2016) and, more recently, an AHP-RIM combination allowed to classify near-Earth Asteroids impact dates (Sánchez-Lozano et al., 2019).

MCDM methodologies such as the TOPSIS, VIKOR or ELECTRE method, classify alternatives according to whether we use benefit or cost criteria, that is, if they must be maximized or minimized. On the other hand, there are decision problems in which the best alternative for one or more criteria is not the one that achieves the highest value or the lowest value for those criteria, but rather they are located within a range of values. When evaluating a series of alternatives based on an ideal solution in such circumstances, RIM method stands out above the rest of the MCDM methodologies (Sánchez-Lozano & Naranjo, 2020). Such method enables to evaluate a large number of alternatives based on a set of criteria through the proximity of these criteria to ideal reference values; these values can also belong to an interval. From the point of view of exoplanet habitability, a clear proof of that are the cases of escape velocity and effective stellar flux criteria. It will not be better the planet having the maximum or minimum value for those criteria over a list of potential candidates, but the one having values for those criteria more similar to the characteristics of our planet. Therefore, this methodology is ideal since it allows to evaluate a high number of exoplanets through a series of criteria based on their proximity to the Earth characteristics.

Furthermore, in order to evaluate exoplanets in terms of their habitability to harbor life, the combination of RIM methodology with techniques such as fuzzy logic (Zadeh, 1965; 1975) allows us to combine quantitative criteria as escape velocity, effective stellar flux, etc., with other criteria that can be interpreted in a subjective way. For example, how worse is a potentially locked planet than an unlocked planet? or a dry-rocky-planet in comparison with another water-rich?, how can the habitability potential be affected if the planet is located at the inner edge of HZ compared to another located at the outer edge? Criteria such as the composition of the planet, the existence of liquid water on its surface or even its tidal lock, can be expressed qualitatively through fuzzy numbers. The combination of RIM method with techniques such as fuzzy logic, provides a very useful way of dealing with this type of exoplanetary habitability analysis. Therefore, we have chosen RIM method and its fuzzy version (Cables et al., 2018) to rank exoplanets, based on a set of criteria, in terms of their similarity to the only ideal alternative that we really know, The Earth (Fig. 1).

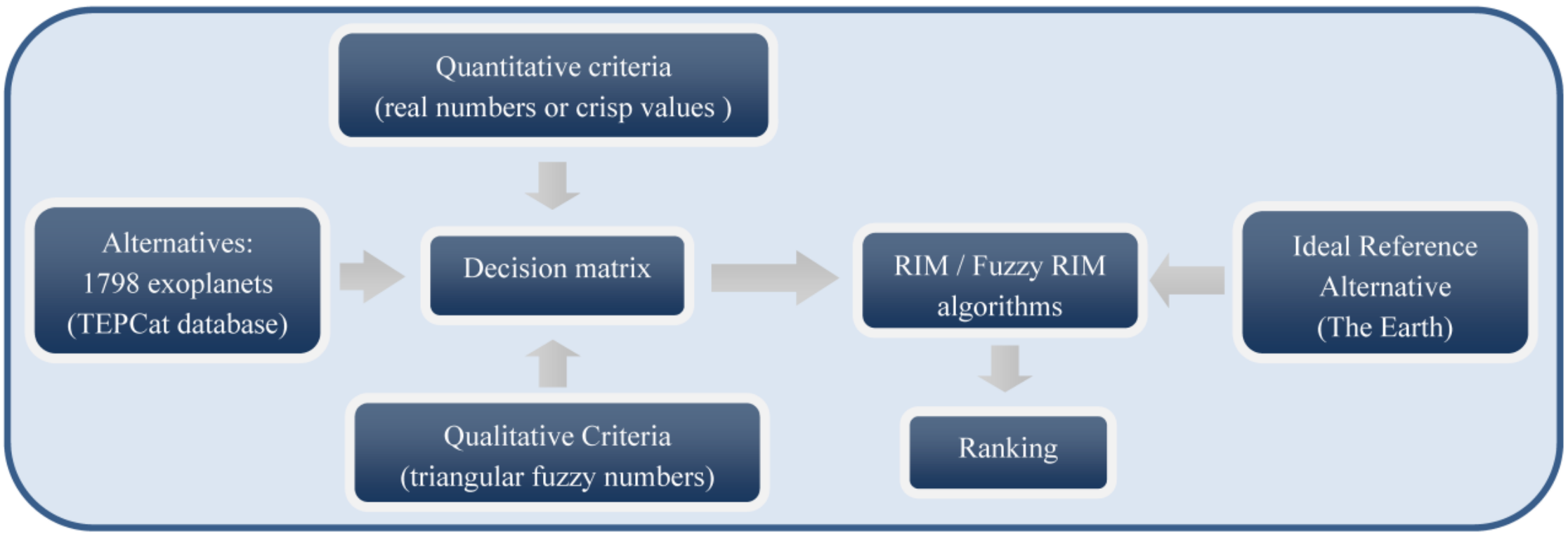

**Fig. 1:** Process scheme

This paper is divided into four sections: Section 2 describes the MCDM methodology used for the proposed decision problem, i.e., the RIM approach and its fuzzy version (FRIM); in Section 3, the description of the selected alternatives to address this case study, and the criteria which have influence in the decision problem, are presented, analyzed and discussed; and finally, Section 4 contains the main conclusions of this study.

## 2. Methodology

### 2.1. Fuzzy sets

Sometimes, it is not easy to quantitatively express qualitative concepts as diverse as the beauty of a landscape, the maturity of a person, or even, from an astrophysical point of view, the habitability potential of a dry rock planet versus a rich in water planet, or a potentially locked planet versus a tidally locked planet. The fuzzy set theory (Zadeh, 1965; 1975) deals with this type of uncertainty. Such theory is a set of concepts and techniques that provides, somehow, a mathematical precision to the human thought process, which is generally imprecise, vague, and ambiguous in the standards of classical mathematics. This fuzzyness is a type of imprecision that can be associated with sets in which there is no abrupt transition between membership or not. If there are criteria that are more appropriate to represent in a qualitative context rather than through the use of real numbers or crisp values (Zadeh, 1975), then the description of a membership function can be further extended to allow for using linguistic labels and variables. Any function of this type is capable of being defined as a membership function associated with fuzzy sets. Its theoretical definition is as follows (Zadeh, 1965):

A fuzzy set $A$ in the space of objects denoted by $X$, where $x$ is a generic element of $X$, is characterized through a membership function $f_A(x)$ which associates with each object in $X$ a real number in the range [0,1]. In this way, the value of $f_A(x)$ at $x$ stands for the grade of membership of $x$ in $A$. Therefore, a high grade of membership of $x$ in $A$ means that the $f_A(X)$ function is close to unity. As a consequence, for every fuzzy set, a membership function must be defined.

Although there are different types of membership functions such as Gaussian function, trapezoidal function, etc., the triangular membership function defined by triangular fuzzy numbers is widely extended thanks to its simplicity and easy handling. This function is specified by three parameters {a, b, c} and defined as follows:

$$\textbf{triangle}\,(\boldsymbol{x};\boldsymbol{a},\boldsymbol{b},\boldsymbol{c}) = \begin{cases} \mathbf{0}, & \boldsymbol{x} \leq \boldsymbol{a} \\ \frac{x-a}{b-a}, & \boldsymbol{a} \leq \boldsymbol{x} \leq \boldsymbol{b} \\ \frac{c-x}{c-b}, & \boldsymbol{b} \leq \boldsymbol{x} \leq \boldsymbol{c} \\ \mathbf{0}, & \boldsymbol{c} \leq \boldsymbol{x} \end{cases}$$

(1)

Such parameters {a, b, c} define the x coordinates of the three vertices of the triangular membership function. Its basic operations are detailed in (Dubois & Prade, 1980; Laarhoven & Pedrycz, 1983; Klir & Yuan, 1995; Triantaphyllou, 2000).

When it is necessary to deal with the problem of ranking $A_i$ ($i=1,\ldots m$) alternatives based on $C_i$ ($i=1,\ldots m$) criteria where qualitative and quantitative criteria coexist, the difficulty of assigning numbers to alternatives in terms of such criteria appears. In this case, MCDM methodologies using real or crisp values may not be directly applicable and we must move to fuzzy versions of these MCDM methodologies and combine criteria of differing natures (qualitative and quantitative). Finally, in order to distinguish the criteria represented by fuzzy numbers from their crisp version counterparts, the former are expressed as $\widetilde{\boldsymbol{C}_i}$.

## 2.2. The Reference Ideal Method (RIM) and its fuzzy version (Fuzzy RIM)

The RIM technique (Cables et al, 2016) enables to evaluate a group of alternatives not only taking into account all the criteria that influence the decision problem simultaneously, but also based on an ideal alternative which can be fictitious or real, the so-called ideal reference alternative. In order to do so, this MCDM calculates, for each alternative, the distances between the value of each criterion and its respective ideal reference value, which may even be a range of values.

The stages of the RIM technique can be summarized as:

**Stage 1.** Defining the study scenario. To do so, the range ($t_j$) which belongs to a given domain $D$, the ideal reference ($s_j$) and the weight ($w_j$) for each specific criterion are defined.

**Stage 2.** Obtaining the decision matrix $X$ composed by $m$ alternatives and $n$ criteria so that $X = (x_{ij})_{i,j}$ for $i=1,\ldots,m$ and $j=1,\ldots,n$.

**Stage 3a. Quantitative Criteria.** Normalizing the decision matrix $X$ through the ideal reference. Therefore, the normalized matrix $Y=(f(x_{ij},t_j,s_j))_{i,j}$ is defined via the piecewise function $f: x \oplus [A,B] \oplus [C,D] \rightarrow [0,1]$ as follows:

$$\boldsymbol{f}(\boldsymbol{x},[\boldsymbol{A},\boldsymbol{B}],[\boldsymbol{C},\boldsymbol{D}]) = \begin{cases} \mathbf{1} & \boldsymbol{if} \quad \boldsymbol{x} \in [\boldsymbol{C},\boldsymbol{D}] \\ \mathbf{1} - \frac{\boldsymbol{d}(\boldsymbol{x},[\boldsymbol{C},\boldsymbol{D}])}{|\boldsymbol{A}-\boldsymbol{C}|} & \boldsymbol{if} \quad \boldsymbol{x} \in [\boldsymbol{A},\boldsymbol{C}] \ \boldsymbol{with}\, \boldsymbol{A} \neq \boldsymbol{C} \\ \mathbf{1} - \frac{\boldsymbol{d}(\boldsymbol{x},[\boldsymbol{C},\boldsymbol{D}])}{|\boldsymbol{D}-\boldsymbol{B}|} & \boldsymbol{if} \quad \boldsymbol{x} \in [\boldsymbol{D},\boldsymbol{B}] \ \boldsymbol{with}\, \boldsymbol{D} \neq \boldsymbol{B} \end{cases} \quad (2)$$

Where for each value criterion $x$: $[\boldsymbol{A},\boldsymbol{B}]$ is the universe of discourse, and $[\boldsymbol{C},\boldsymbol{D}]$ is the range of the ideal reference, so, $\in [\boldsymbol{A},\boldsymbol{B}]$ and $[\boldsymbol{C},\boldsymbol{D}] \subset [\boldsymbol{A},\boldsymbol{B}]$ . The distance to the ideal reference is obtained by using the following expression:

$$\boldsymbol{d}(\boldsymbol{x},[\boldsymbol{C},\boldsymbol{D}]) = \boldsymbol{min}\{|\boldsymbol{x}-\boldsymbol{C}|,|\boldsymbol{x}-\boldsymbol{D}|\} \quad (3)$$

**Stage 3b. Qualitative Criteria**. In case of having qualitative values defined through triangular fuzzy numbers, it is necessary to apply the fuzzy version of the RIM technique, so-called Fuzzy RIM (Cables et al., 2018). In such case, the distance between two fuzzy numbers $(\tilde{X}_{ij}, \tilde{D}_{ij})$ is obtained as follows:

$$\boldsymbol{dist}(\tilde{\boldsymbol{X}}_{ij}, \tilde{\boldsymbol{D}}_{ij}) = \sqrt{\frac{1}{3}((\boldsymbol{x}_1 - \boldsymbol{d}_1)^2 + (\boldsymbol{x}_2 - \boldsymbol{d}_2)^2 + (\boldsymbol{x}_3 - \boldsymbol{d}_3)^2)} \tag{4}$$

Since a triangular fuzzy number is defined by a membership function represented by three points, the upper and lower extremes and the modal value, such distance corresponds to that between the vertices of two triangular membership functions located in a plane.

In the same manner, the distance to the fuzzy ideal reference interval $\boldsymbol{I}\tilde{\boldsymbol{R}}_j = [\tilde{\boldsymbol{C}}_j, \tilde{\boldsymbol{D}}_j]$ is calculated by using the following expression:

$$\boldsymbol{d}^*_{\boldsymbol{min}}(\tilde{\boldsymbol{X}}_{ij}, [\boldsymbol{I}\tilde{\boldsymbol{R}}_j]) = \boldsymbol{min}\left(\boldsymbol{dist}(\tilde{\boldsymbol{X}}_{ij}, \tilde{\boldsymbol{C}}_j), \boldsymbol{dist}(\tilde{\boldsymbol{X}}_{ij}, \tilde{\boldsymbol{D}}_j)\right) \tag{5}$$

where $\tilde{\boldsymbol{X}}, \tilde{\boldsymbol{C}}$ and $\tilde{\boldsymbol{D}}$ are triangular fuzzy numbers.

Likewise, the expression to normalize the decision matrix must be modified as follow:

$$\boldsymbol{f}^*(\tilde{\boldsymbol{X}}_{ij}, [\tilde{\boldsymbol{R}}_j], [\boldsymbol{I}\tilde{\boldsymbol{R}}_j]) =$$

$$\begin{cases} \mathbf{1} & \boldsymbol{if} \quad \tilde{\boldsymbol{X}}_{ij} \in [\boldsymbol{I}\tilde{\boldsymbol{R}}_j] \\ \mathbf{1} - \frac{\boldsymbol{d}^*_{\boldsymbol{min}}(\tilde{\boldsymbol{X}}_{ij}, [\boldsymbol{I}\tilde{\boldsymbol{R}}_j])}{\boldsymbol{dist}(\tilde{\boldsymbol{A}}_j, \tilde{\boldsymbol{C}}_j)} & \boldsymbol{if} \tilde{\boldsymbol{X}}_{ij} \in [\tilde{\boldsymbol{A}}_j, \tilde{\boldsymbol{C}}_j] \wedge \tilde{\boldsymbol{X}}_{ij} \notin [\boldsymbol{I}\tilde{\boldsymbol{R}}_j] \wedge \boldsymbol{dist}(\tilde{\boldsymbol{A}}_j, \tilde{\boldsymbol{C}}_j) \neq \mathbf{0} \\ \mathbf{1} - \frac{\boldsymbol{d}^*_{\boldsymbol{min}}(\tilde{\boldsymbol{X}}_{ij}, [\boldsymbol{I}\tilde{\boldsymbol{R}}_j])}{\boldsymbol{dist}(\tilde{\boldsymbol{D}}_j, \tilde{\boldsymbol{B}}_j)} & \boldsymbol{if}\ \tilde{\boldsymbol{X}}_{ij} \in [\tilde{\boldsymbol{D}}_j, \tilde{\boldsymbol{B}}_j] \wedge \tilde{\boldsymbol{X}}_{ij} \notin [\boldsymbol{I}\tilde{\boldsymbol{R}}_j] \wedge \boldsymbol{dist}(\tilde{\boldsymbol{D}}_j, \tilde{\boldsymbol{B}}_j) \neq \mathbf{0} \\ \mathbf{0} & \boldsymbol{in\ other\ case} \end{cases} \tag{6}$$

Where $\tilde{\boldsymbol{R}}_j = [\tilde{\boldsymbol{A}}_j, \tilde{\boldsymbol{B}}_j]$ represents the universe of discourse in fuzzy numbers.

**Stage 4.** Weighting the normalized matrix taken into consideration the criteria weights. The weighted normalized matrix takes the following form:

$$\boldsymbol{Y'} = \boldsymbol{Y} \otimes \boldsymbol{W} = (\boldsymbol{y_{ij}} \cdot \boldsymbol{wj})_{\boldsymbol{ij}} \text{ for } i=1,\ldots,m \text{ and } j=1,\ldots,n. \quad (7)$$

**Stage 5.** Calculating the distance, for each alternative, to the normalized ideal reference:

$$\boldsymbol{I_i^+} = \left\{\sum_{j=1}^{n} \left(\boldsymbol{y'_{ij}} - \boldsymbol{w_j}\right)^2\right\}^{\frac{1}{2}} \text{ and } \boldsymbol{I_i^-} = \left\{\sum_{j=1}^{n} \left(\boldsymbol{y'_{ij}}\right)^2\right\}^{\frac{1}{2}} \text{ for } i=1,\ldots,m \text{ and } j=1,\ldots, n. \quad (8)$$

**Stage 6.** Determining a relative index of each alternative by using the following expression:

$$\boldsymbol{R_i} = \frac{\boldsymbol{I_i^-}}{\boldsymbol{I_i^+} + \boldsymbol{I_i^-}}, \boldsymbol{i} = \boldsymbol{1}, \ldots, \boldsymbol{m}, \text{ where } \boldsymbol{R_i} \in (\boldsymbol{0}, \boldsymbol{1}). \quad (9)$$

**Stage 7.** Ranking all the alternatives in descending order based on their relative indexes. Therefore, those alternatives located in the first positions of the ranking constitute the best solutions.

The RIM approach and its fuzzy version (Fuzzy RIM) will be used in this study to evaluate alternatives, i.e., to rank exoplanets, based on a set of criteria, in terms of their potential habitability according to their similarity to the Earth.

## 3. The decision problem: Analysis of the potential habitability of exoplanets based on their similarity to Earth

In the effort of searching for live outside our solar system, the first steps have been already done and nowadays thousands of exoplanets are known, some of them Super- and Exo-Earths potentially habitable. In the near future, mainly thanks to current and future space missions such as TESS and Plato2.0, this database is expected to exploit. Therefore, we are ready for the next steps: searching for biomarkers. Near future facilities with such capacities, as it is the case of JWST, won't be able to perform large surveys. Therefore, ranking known exoplanets to prioritize them is a must if we want to increase the efficiency of these instruments.

Therefore, the main objective of this work is to provide this ranking taking the advantage of recently developed fuzzy RIM approach. This task will be done based on a set of criteria which have influence in the evaluation process in terms of their similarity to the Earth. To obtain the physical characteristics of known exoplanets and their hosting stars for training purposes we have used the TEPCat database (Southworth, 2020) which contains a catalogue of the physical properties of transiting planetary systems. The decision of using this dataset is based on our interest in using the most reliable and homogeneous data possible despite the fact that in the literature there are more comprehensive datasets in terms of the number of exoplanets included. For this study we have used the 1798 exoplanets of this database with hosting stars with effective temperatures lower than 9000 K, and with masses undoubtedly in the range of planetary masses. The mentioned database contains a summary of physical properties commonly provided by the planet discovery projects such as orbital period, eccentricity, semimajor axis, stellar properties ($T_{eff}$, [Fe/H], mass, radius, etc.) and planetary properties (mass, radius, gravity, etc.) for all transiting extrasolar planetary systems which have been analyzed in-depth.

### 3.1. Criteria definition

In order to analyze the main parameters which have influence in the exoplanetary habitability, the current version of the Statistical-likelihood Exo-Planetary Habitability Index (SEPHI, Rodríguez-Mozos & Moya, 2017) has been applied to the TEPCat database. This new version of SEPHI differs from its initial version in the definition of the HZ and the influence of the eccentricity of the orbit on the average flux received by the planet. Related to the HZ, the influence of the Mass of the planet on its limits (Kopparapu et al., 2014) and the displacement of the inner Edge for M red dwarfs (Kopparapu et al., 2013) have been included.

In this way, exoplanets parameters such as internal composition (core, mantle and ice mass fraction), capacity for retaining an atmosphere (escape velocity), incident energy (effective stellar flux), tidal locking (exoplanet magnetic moment and stellar tidal locking zones), type of planet (telluric, ice, ice giant and gas giant) and the potential presence of liquid water on its surface (habitable zone and its boundaries) constitute the criteria having influence in the assessment of alternatives, from the perspective of the exoplanetary habitability and their similarity to the Earth (Fig. 2).

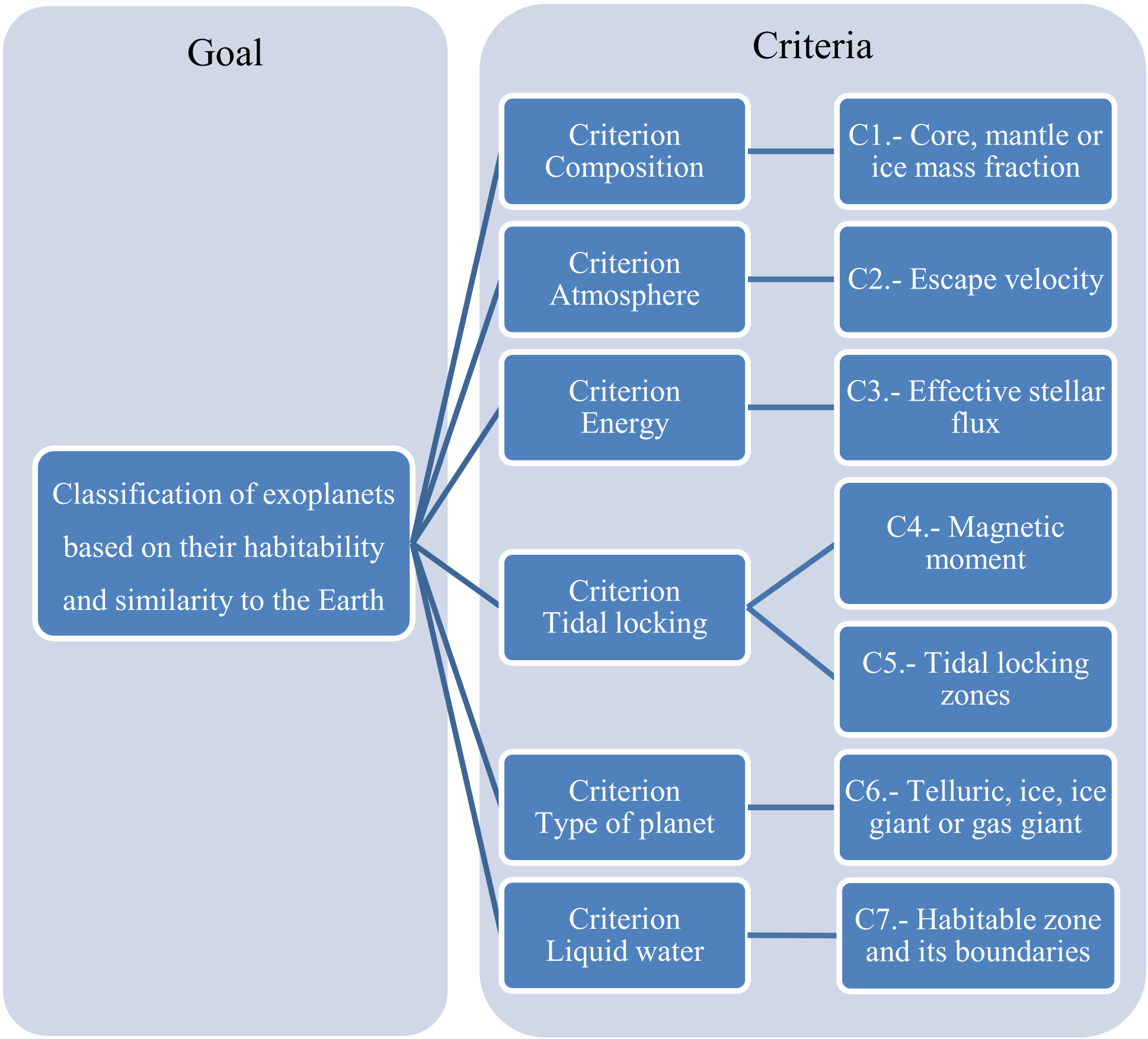

**Fig. 2:** Criterion tree resulting from the problem structuring phase

### 3.1.1. Criterion Composition. $C_1$.- Mass fractions of Fe, $MgSiO_3$ or $H_2O$

To define the internal structure of rocky planets, both dry and water rich, we will use the mass-radius diagram of (Zeng & Sasselov, 2013), and the mass fractions of the total mass of the planet corresponding to the core (CMF), Mantle (MMF), and Ice cap (IMF). However to explain the internal structure of Giants planets, has been used the work of (Fortney et al., 2007), and the mass fractions of the total mass of the planet corresponding to the Nucleus (NMF) and Envelope (EMF).

The obtained values of CMF, MMF and IMF allow us to define the composition criterion ($C_1$). Since the group of exoplanets that constitute the alternatives of our decision problem covers different types, in our study we will consider CMF the driving value to study the composition of telluric planets, and the IMF value that for study the composition of the ice planets. The probability of giant planets being habitable is barely null, and therefore it is not necessary to analyze their internal structure in depth.

Therefore, in this study the following types of exoplanets will be considered:

- **Telluric or Dry Rocky exoplanets.** Those exoplanets that, for a given mass, have a radius equal or less that the radius of a planet with CMF=0 in the mass-radius diagram. They are mainly iron and silicate worlds where the presence of water is residual, as on Earth.
- **Ice or Water Rich Rocky exoplanets.** They are those exoplanets that, for a given mass, have a radius greater than the radius of a planet with CMF=0 and equal, or lower, than the radius corresponding to a planet with IMF= 1. That is, from the radius corresponding to CMF=0, the decrease in density cannot be justified with the increased presence of silicates and the presence of a lighter element such as water is necessary. If an exoplanet of this type is in HZ, we call it “Ocean planet”.

- **Ice Giants.** They are worlds with a nucleus mainly composed by Ice and Silicates, and a gas envelope in the outer zone mainly of Hydrogen and Helium, where the planetary nucleus predominates. They are those planets that, for a given mass, have a radius greater than a planet with IMF=1 and lower, or equal, to the radius corresponding to a planet with a 50% H-He. That is, from the radius corresponding to IMF=1, the decrease in density cannot be justified with the increased presence of water and the presence of a lighter element such Hydrogen and Helium is necessary. If an exoplanet of this type has a normalized radius less to 3, we call it "Mini Neptune".
- **Gas Giants**. They are worlds of Ice and Silicates in the nucleus, and a gas envelope in the outer zone of Hydrogen and Helium, where the envelope predominates. They are those planets that, for a given mass, have a radius greater than the corresponding to planets with a 50% H-He.

The dry rocky planets will be those whose water mass is negligible compared to the total mass of the planet, and therefore its internal structure can be explained only by two components, iron in the core and silicates in the mantle. This would be the case for the inner planets of the solar system such as Mercury, Venus, Earth and Mars. When there are only two basic constituents, for a given planetary mass and radius, the theoretical models provide a unique solution for its internal structure (Suissa et al., 2018; Zeng & Sasselov, 2013).

However, water-rich rocky exoplanets need three basic components to explain their internal structure: iron in the core, silicates in the mantle, and water usually forming an outer layer of ice. In this case, for a given mass and radius, the theoretical models provide infinite internal structure solutions all compatible with those mass and radius values (Suissa et al., 2018; Zeng & Sasselov, 2013).

### 3.1.2. Criterion Atmosphere. $C_2$.- Escape velocity ($v_e$)

According to the study carried out by Frederick & Lubin (1988) on the effect of ultraviolet radiation on the Earth-Atmosphere system, in clear sky conditions and at short wavelengths (less than 304 nm), 90% of the incident radiation is absorbed by the atmosphere. Such percentage of radiation falls to 50% between 313 and 314 nm. That study is a clear example of the protective shield that the atmosphere offers. Its long-term stability is one of the widely accepted requirements as unquestionable needs for the resurgence and evolution of life (Lammer et al., 2010).

Among its causes, a high density atmosphere stands out above the rest since the Earth's atmosphere contains numerous traces of gases such as $CH_4$, $N_2O$, $CH_3Cl$, COS, etc. In fact, some of these gases ($CH_4$ and $N_2O$ for example) influence climate change by contributing to the atmospheric greenhouse effect (Kasting & Siefert, 2002), others such as $CO_2$, (one of the main greenhouse gases) can protect atmospheres and water reserves against evaporation during periods of extreme ultraviolet radiation from its star (Lammer et al., 2010) while $O_2$ combined with a moderate temperature (300 ± 50 K) is a strong indicator of life based on photosynthesis on distant planets (Leger et al., 1993).

Planets present different ways of losing their atmosphere such as thermal and non-thermal escape, and erosion by stellar wind, to mention just a few of them. Among these forms, gravitational escape is a first approximation allowing us to predict whether a planet is capable of retaining its atmosphere. Low mass planets do not exert a high resistance to the atmosphere erosion so that light particles in their atmosphere can present enough energy to allow them to leave it. This atmosphere erosion resistance can be parametrized through the normalized escape velocity ($v_e$):

$$v_e = \sqrt{\frac{M_p}{R_p}} \tag{10}$$

Where, Mp and Rp are the normalized planetary mass and the radius respectively.

In our study, the escape velocity constitutes the atmosphere criterion ($C_2$).

### 3.1.3. Criterion Energy. $C_3$.- Normalized effective stellar flux ($S_{eff}$)

In the Earth, the effective stellar flux impacting its surface plays a major role on the main biological processes. This energy source, together with carbon dioxide and water, allow photosynthesis. The oxygen generated by photosynthesis provides an abundant source of energy for microorganisms. In this way, these organisms could find a reliable and durable energy source that would allow them to survive, multiply and evolve on planetary time scales (Wolstencroft & Raven, 2002).

In order to generate a $O_2$ molecule via photosynthesis process, the number of photons required is closely related to the wavelength since the higher the long wavelength limit for photosynthesis, the greater the number of photons necessary to reduce a $CO_2$ molecule to carbohydrates or to develop an $O_2$ molecule (Heath et al., 1999). In particular, Photosynthetically Active Radiation (PAR) is mainly in the wavelength range of 400 to 720 nm and in fact, the solar flux on the Earth's surface reaches a maximum at 685 nm (Gebauer, 2003; Gebauer et al., 2017).

The stellar flux a planet receives is inversely proportional to the square of the distance of that planet's orbit from its star, specifically the semi-major axis of its orbit. Therefore, the energy flux provided by the star in the exoplanet's orbit can be calculated through the normalized effective flux incident on a planet ($S_{eff}$) in circular orbit around its star. In this work, such parameter constitutes the energy criterion ($C_3$):

$$S_{eff} = \frac{L_*}{a^2} \tag{11}$$

Where, $L_*$ is the normalized luminosity of the star with respect to the Sun and $a$ the semi-major orbital axis.

### 3.1.4. Criterion Tidal locking. $C_4$.- Magnetic moment and $C_5$.- Tidal locking zones (qualitative)

The tidal locking phenomenon is generated through tidal friction processes, causing warming, that is the reason why this phenomenon is also called tidal warming. The energy generated by these friction processes can be dissipated in the form of heat both on the surface of the oceans and inside the celestial bodies. When a planet orbits close to its main star, it is forced to rotate synchronously due to these tidal forces (Huang, 1960), and therefore to have a permanently illuminated hemisphere in case of orbital resonance 1:1, as it is the case of the Earth – Moon system.

These conditions are not ideal from the point of view of habitability since such planets receive a very unequal stellar heating. On the one hand, the side that faces its star is a day side and may be hot enough to hold liquid water. On the other hand, on the dark and nocturnal side, the night could be so cold that any gas could condense (Hu & Yang, 2014). These characteristics could generate a phenomenon called atmospheric collapse which has been postulated as the main reason against searching for habitable planets around M stars (Joshi et al. 1997).

Furthermore, the planets affected by this tidal lock have a weak magnetic field due to their low angular frequency. This magnetic field acts as a protective shield against cosmic and stellar radiation (Grießmeier et al., 2005; Rekola, 2009; Rodriguez-Mozos and Moya, 2019). In SEPHI, the exoplanet magnetic moment is normalized taking the Earth as reference through the following expression:

$$M_n = \alpha\, \rho_{0n}^{1/2}\, r_{0n}^{10/3}\, F_n^{1/3} \qquad (12)$$

Where $\rho_{0n}$, $r_{0n}$ and $F_n$ represent the normalized core average density, the normalized radius and the normalized convective buoyancy flux, respectively (Rodriguez-Mozos & Moya, 2017). $\alpha$ is a reduction factor depending on the planet's magnetic regime. For dipolar regime $\alpha$=1 while for multipolar magnetic regime $\alpha$=0.15.

This magnetic moment constitutes the first criterion ($C_4$) of the tidal locking criterion set.

The different tidal locking regimes for an exoplanet similar to The Earth were initially defined by Grießmeier et al. (2005) and improved afterwards by Grießmeier et al. (2009). They graphically showed three different regimes (locked, potentially locked or unlocked) as a function of the stellar host spectral type or its mass (M*) and its orbital radius (d). These relationships and mathematical expressions can be seen in detail in Grießmeier et al. (2009). In order to answer questions like: how worse is a potentially locked planet than an unlocked planet?, and a tidally locked planet?, we define a new criterion named $C_5$.-Tidal locking zones (tidally locked, potentially locked and unlocked) which can be expressed qualitatively through fuzzy numbers.

### 3.1.5. Criterion Type of planet. $C_6$.- Substrate (qualitative)

As a consequence of plate tectonics, solid planetary nucleus favors habitability (Kasting, 1993; Sundquist, 1993). That is due to the fact that plate tectonics allow recycling of $CO_2$, a fundamental process to protect planet climate from stellar luminosity (Kasting & Catling, 2003).

In order to define this criterion, it is necessary to determine the exoplanet type from the point of view of its internal structure. For doing so, the planetary mass and radius is required. Through different studies of mass-radius relationships for solid exoplanets, it is possible to distinguish gaseous mini-Neptunes, water worlds, and rocky Super-Earths (Valencia et al., 2006; Seager et al., 2007; Dressing et al., 2015). More recently, Chen & Kipping (2017) extended this classification distinguishing between terrestrial, Neptunian (ice giant) and Jovian (gas giant) worlds through the following expression:

$$\frac{R_*}{R_\oplus} = C\left(\frac{M_*}{M_\oplus}\right)^S \qquad (13)$$

Where R* and M* are the values of the planetary radius and mass, $\boldsymbol{R}_\oplus$ and $\boldsymbol{M}_\oplus$ correspond to the terrestrial values, $C$ and $S$ are parameters obtained through a power law.

Although from the point of view of exoplanetary habitability ice and gas giant worlds do not offer excessive probabilities for supporting life, such possibility cannot totally be excluded (Lammer et al., 2009; Schulze-Makuch et al., 2011). We will take this potential difference into consideration in this study.

Therefore, how worse is a dry rock planet than a rich in water planet? or, a Neptunian one in comparison with a Jovian planet?. Parameters such as the planet typology (telluric, ice, ice giants and gas giants worlds) have a strong qualitative character and can be therefore expressed through fuzzy logic via triangular fuzzy numbers. That typology constitutes the $C_6$ criterion (Substrate).

### 3.1.6. Criterion Liquid water. $C_7$.- Habitable zones and its boundaries (qualitative)

Liquid water on the surface of a planet is one of the critical conditions from the perspective of exoplanetary habitability (Lammer et al., 2009; Madhusudhan et al., 2016). Such characteristic is closely linked to the concept of Habitable Zone (HZ) which has been deeply studied and analyzed over years, slightly varying the climatic restrictions that define its boundaries (Hart, 1979; Kasting, 1988; Kasting et al., 1993; Selsis et al., 2007; Kaltenegger & Sasselov, 2011, Bin et al., 2018).

Based on the boundaries of the HZ defined by Kopparapu et al. (2013; 2014) and Selsis et al. (2007), SEPHI estimates the following zones according to the orbital semi-mayor axis $a$ (for more details, see Rodríguez-Mozos & Moya, 2017):

- Hot zone: Probability of water in gaseous form
- Inner Transition Zone / Outer Transition Zone: Non-zero probability that a planet has liquid water.
- Green Zone: Maximum probability (unit value) that a planet has liquid water
- Cold Zone: Probability of water in solid form

Once again, we can ask questions like: how can the habitability of a planet be affected if it is located at the Inner Transition Zone compared to another located at the Outer edge?. In order to try to answer questions of this type, we define the criterion $C_7$ (habitable zones and its boundaries) in a subjective way based on the location of a given exoplanet in one of the previously mentioned zones.

This criterion is strongly correlated with criterion $C_3$, but integrates additional information and assumptions. Therefore, in order to avoid taking into account the shared information twice in the final result, we have weighted these two criteria with a factor one half each.

### 3.2. Obtaining assessments of the alternatives

Once the criteria have been defined, we can apply the Fuzzy RIM methodology to rank as much alternatives as we want. In our case, we will study a total of 1798 exoplanets with quite reliable determinations of their main physical, orbital, and stellar host characteristics. These values have been taken from TEPCat database (Southworth, 2020) and they are the basic ingredients for our study. The most complex ingredients for achieving the criteria have been obtained using the formalism described by Rodríguez-Mozos & Moya (2017, 2019), in the context of the SEPHI index (this is what we call "SEPHI formalism"). Finally, each one of the criteria must be valued based on a reference ideal alternative, The Earth in this case.

Criteria $C_1$, $C_2$, $C_3$ and $C_4$ are determined numerically via SEPHI formalism. An example for ten exoplanets is shown in Table 1.

| | Criterion Composition | Criterion Atmosphere | Criterion Energy | Criterion Tidal locking |
|---|---|---|---|---|
| | $C_1$.-CMF, MMF or IMF (%) | $C_2$.- Escape velocity ($v_e$) | $C_3$.-Effective stellar flux ($S_{eff}$) | $C_4$.-Magnetic moment ($M$) |
| Kepler-442b | 0.21 | 1.35 | 0.60 | 1.93 |
| Kepler-062f | 0.09 | 1.37 | 0.50 | 1.88 |
| Kepler-441b | 0.26 | 1.43 | 0.20 | 2.20 |
| Kepler-062e | 0.28 | 1.43 | 1.41 | 1.55 |
| LHS_1140b | 0.36 | 2.00 | 0.48 | 0.26 |
| Kepler-069c | 0.39 | 1.43 | 2.05 | 2.37 |
| Kepler-186f | 0.21 | 1.18 | 0.35 | 0.10 |
| Kepler-439b | 0.41 | 1.65 | 1.81 | 3.41 |
| Kepler-344c | 0.00 | 1.80 | 4.26 | 13.46 |
| K2-009b | 0.27 | 1.42 | 1.15 | 0.24 |

**Table 1:** Example of decision matrix composed by ten alternatives and four criteria

Likewise, the rest of the criteria ($C_5$, $C_6$ and $C_7$), for each alternative, are obtained via SEPHI formalism. However, in that case, such criteria do not present numerical values but qualitative forms. Again, an example for the exoplanets mentioned above is shown in Table 2.

| | Criterion Tidal locking | Criterion Type of planet | Criterion Liquid water |
|---|---|---|---|
| | $C_5$.- Tidal locking zones | $C_6$.- Substrate | $C_7$.- Habitable zones and its boundaries |
| Kepler-442b | Potentially Locked | Telluric | Green |
| Kepler-062f | Unlocked | Ice | Green |
| Kepler-441b | Unlocked | Ice | Outer edge |
| Kepler-062e | Potentially Locked | Ice | Inner edge |
| LHS_1140b | Tidally Locked | Telluric | Green |
| Kepler-069c | Unlocked | Ice | Inner edge |
| Kepler-186f | Potentially Locked | Telluric | Green |
| Kepler-439b | Potentially Locked | Ice | Inner edge |
| Kepler-344c | Potentially Locked | Ice giant | Inner edge |
| K2-009b | Tidally Locked | Ice | Inner edge |

**Table 2:** Example of decision matrix composed by ten alternatives and three subjective criteria

In order to value such qualitative criteria, we use linguistic labels associated to triangular fuzzy numbers (Table 3). Criterion $C_5$ divides the tidal locking zones into linguistic labels from "Tidally Locked" to "Unlocked". A "Tidally Locked" exoplanet is very close to its star and rotates synchronously with its orbits. Due to this fact, not only stellar winds impact heavily the exoplanet, furthermore, such it presents difficulties to generate a magnetic field to protect its surface from them. On the contrary, an "Unlocked" exoplanet is not that close to its star and it rotates freely. Hence, it is more protected from stellar winds on the one hand, and it has more options to generate a magnetic field, like The Earth, on the other. Finally, a "Potentially Locked" exoplanet is not blocked at the beginning so it could develop life that later adapts to the slow braking of its rotation over time. Therefore, from the point of view of the habitability and similarity to The Earth, an "Unlocked" exoplanet should present the highest fuzzy valuation (Table 3 and Figure 3).

| Valuations of exoplanets based on their habitability and similarity to the Earth | |
|---|---|
| Tidal locking zones ($C_5$) | |
| Linguistic labels | Triangular fuzzy numbers |
| Tidally Locked | (0, 1, 3) |
| Potentially Locked | (3, 5, 7) |
| Unlocked | (7, 9, 10) |

**Table 3:** Triangular fuzzy numbersof criteria $C_5$ - Tidal locking zones

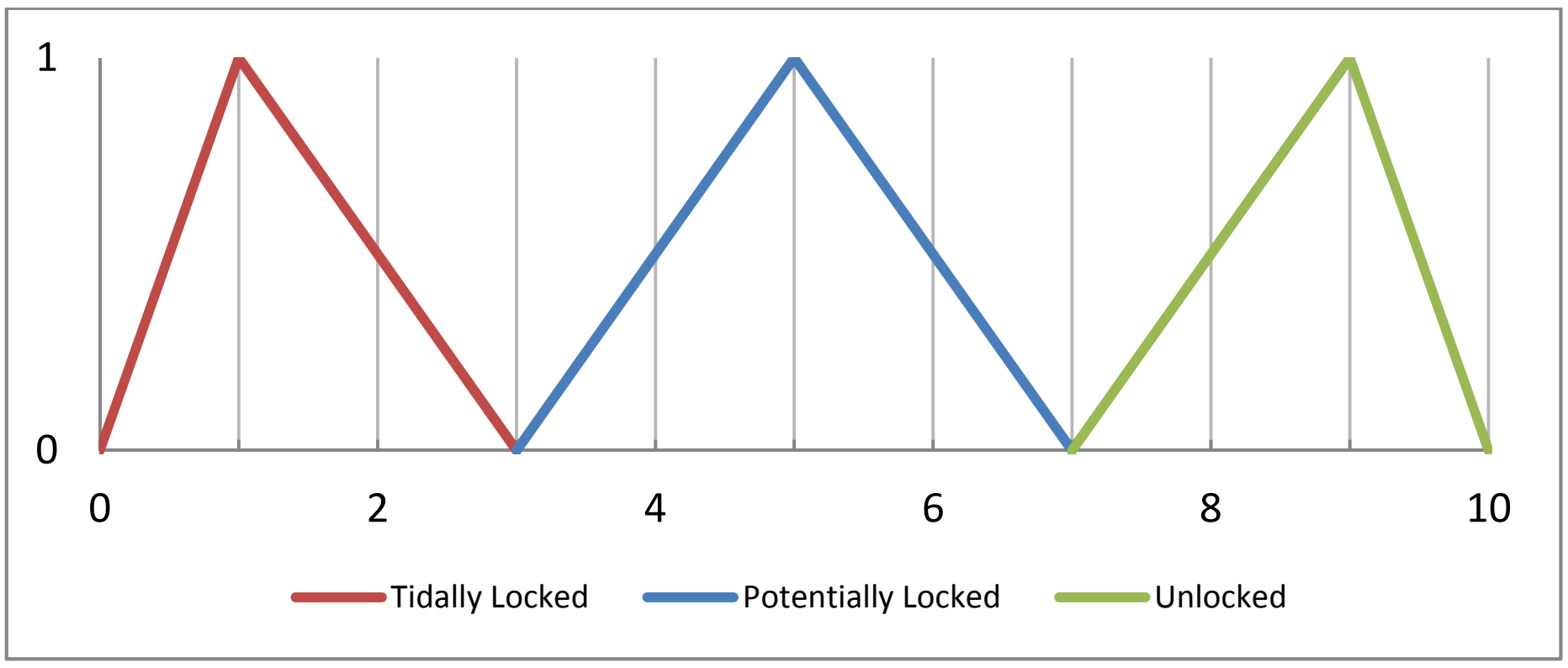

**Fig. 3:** Graphical Representation of the linguistic labels of criteria $C_5$ - Tidal locking zones

The same procedure is applied to the substrate ($C_6$) and liquid water ($C_7$) criteria. In these cases, four triangular fuzzy numbers are defined to distinguish between Jovian and Neptunian worlds for the substrate criterion, and hot or cold planets and those located at the outer edge of HZ for the liquid water criterion (Table 4). These fuzzy scales are based on the characteristics of each criterion. In the case of the Substrate criterion ($C_6$), the typology of a planet cannot be a combination of different typologies. For example, a telluric planet cannot coexist with an ice planet at the same time. On the other hand, the boundaries of the habitable zones ($C_7$) are not so well defined and the fuzzy logic allows us to take into consideration this lack of definition, colliding the triangular membership functions of outer and inner edge.

| Valuations of exoplanets based on their habitability and similarity to the Earth | | | |
|---|---|---|---|
| Substrate ($C_6$) | | Habitable zones and its boundaries ($C_7$) | |
| Planet typology | Triangular fuzzy numbers | Proximity to HZ | Triangular fuzzy numbers |
| Gas giant | (0, 0, 1) | Hot / Cold | (0, 0, 1) |
| Ice giant | (0, 1, 3) | Outer edge | (1, 3, 5) |
| Ice | (3, 5, 7) | Inner edge | (3, 5, 7) |
| Telluric | (7, 9, 10) | Green | (7, 9, 10) |

**Table 4:** Triangular fuzzy numbers of $C_6$ (Substrate) and $C_7$ (Habitable zones and its boundaries) criteria

The fuzzy scales shown (Tables 3 and 4) are based on the standard 7-element fuzzy scale (García-Cascales et al., 2012).

Thus, the corresponding information of the criteria for the 1798 exoplanets generates a decision matrix of alternatives and criteria. Due to the coexistence of quantitative and qualitative criteria, such decision matrix is composed by a mixture of crisp numerical values and triangular fuzzy numbers. In Table 5, the decision matrix for this particular case is shown, where $x_{ij}$ represents the value of the alternative $A_i$ with respect to crisp (real number) criterion $C_j$ and, $\tilde{x}_{ij}$ such value but regarding the fuzzy criterion $\tilde{C}_j$.

| | $C_1$ | $C_2$ | $C_3$ | $C_4$ | $\widetilde{C_5}$ | $\widetilde{C_6}$ | $\widetilde{C_7}$ |
|---|---|---|---|---|---|---|---|
| $A_1$ | $x_{11}$ | $x_{12}$ | $x_{13}$ | $x_{14}$ | $\tilde{x}_{15}$ | $\tilde{x}_{16}$ | $\tilde{x}_{17}$ |
| $A_2$ | $x_{21}$ | $x_{22}$ | $x_{23}$ | $x_{24}$ | $\tilde{x}_{25}$ | $\tilde{x}_{26}$ | $\tilde{x}_{27}$ |
| … | … | … | … | … | … | … | … |
| $A_{1798}$ | $x_{1798\ 1}$ | $x_{1798\ 2}$ | $x_{1798\ 3}$ | $x_{1798\ 4}$ | $\tilde{x}_{1798\ 5}$ | $\tilde{x}_{1798\ 6}$ | $\tilde{x}_{1798\ 7}$ |

**Table 5:** Decision matrix for the assessment of exoplanets based on their habitability

### 3.2.1. The ideal reference

A critical step for the Fuzzy RIM methodology is to define the values (or value ranges) of the ideal reference and the general validity ranges or universe of discourse for every criteria. In this study, universe of discourse $[A, B]$ covers all the existing values for each criterion in the decision matrix. In that case, such range is as follow (Table 6):

| *Range* | $C_1$ *(%)* | $C_2$ *($V_e$)* | $C_3$ *($S_{eff}$)* | $C_4$ *($M$)* | $C_5$ *(fuzzy)* | $C_6$ *(fuzzy)* | $C_7$ *(fuzzy)* |
|---|---|---|---|---|---|---|---|
| *A* | 0.02 | 0.3 | 0.08 | 0.01 | (0, 0, 1) | (3, 3.5, 4) | (1, 1.5, 2) |
| *B* | 1 | 50.65 | 9287.03 | 999999 | (9, 9.5, 10) | (9, 9.5, 10) | (9, 9.5, 10) |

**Table 6:** Values of the range or universe of discourse

On the other hand, since our decision problem consists of the assessment of exoplanets based on their habitability and similarity to the Earth, the range of the reference ideal alternative $[C, D]$ correspond to those that allow to harbor life similar to that of our planet (Table 7).

| Ideal Reference | $C_1$ (%) | $C_2$ ($V_e$) | $C_3$ ($S_{eff}$) | $C_4$ ($M$) | $C_5$ (*fuzzy*) | $C_6$ (*fuzzy*) | $C_7$ (*fuzzy*) |
|---|---|---|---|---|---|---|---|
| *C* | 0.1 | 0.8 | 0.7 | 1 | (6.5, 7, 7.5) | (6.5, 7, 7.5) | (6.5, 7, 7.5) |
| *D* | 0.5 | 3 | 1.05 | 10 | (8.5, 9.5, 10) | (8.5, 9.5, 10) | (8.5, 9.5, 10) |

**Table 7:** Values of the ideal reference alternative

The Core Mass Fraction (CMF) is used to assess the composition criterion ($C_1$) of dry telluric exoplanets and it is related to the presence of a magnetic field protecting life. The minimum time for elementary life to significantly alter a planet's atmosphere (biomarkers) up to detectable levels from another planetary system has been estimated at 2 Gyr (Truitt et al., 2015). Therefore, the planetary dynamo must protect life efficiently, for at least these 2 Gyr, so that it can develop and become detectable. So that, if CMF = 0, its magnetic moment should be null, whereas if CMF = 1 its core would solidify in a very short time ($t_{dyn} \approx 0$) and therefore it would not offer protection either.

The protecting time provided by the dynamo of dry teluric planets decreases when the mass of the planet or its CMF increase (Zuluaga & Bustamante, 2018**)**. For dry rocky planets with a mass of 1.5 $M_{\oplus}$ when CMF > 0.4 the shutdown time of the planetary dynamo is less than 2 Gyr (Zuluaga & Bustamante, 2018). However, telluric planets with small nucleus (CMF = 0.1) can have dynamos for a longer time due to a greater resistance to changing into a convective flow taking into account the higher thickness of the mantle. Nevertheless, they have a weak magnetic protection due to the small size of the nucleus. Therefore, for dry telluric planets we will consider an ideal interval for CMF between [0.1,0.5].

Ice Mass Fraction (IMF) is used to assess the composition criterion ($C_1$) of ice or water-rich planets because, generally, such planets have longer planetary dynamos than dry planets and $t_{dyn}$ is not a determining factor (Zuluaga & Bustamante, 2018). Furthermore, when IMF = 0, that is, we have a completely rocky planet, the magnetic protection of this type of planets would be null. Likewise, planets made up entirely of water (IMF =

1) would not be suitable for the origin of life since there is no a nucleus. On the other hand, it has been considered that in those exoplanets with most of their mass in the form of water (IMF>0.5) the solid substrate would be found at a depth where pressure and temperature would make the origin of life not possible.

The graphical representation of Composition criterion ($C_1$) ideal values for telluric, ice or water-rich planets is shown (Figure 4).

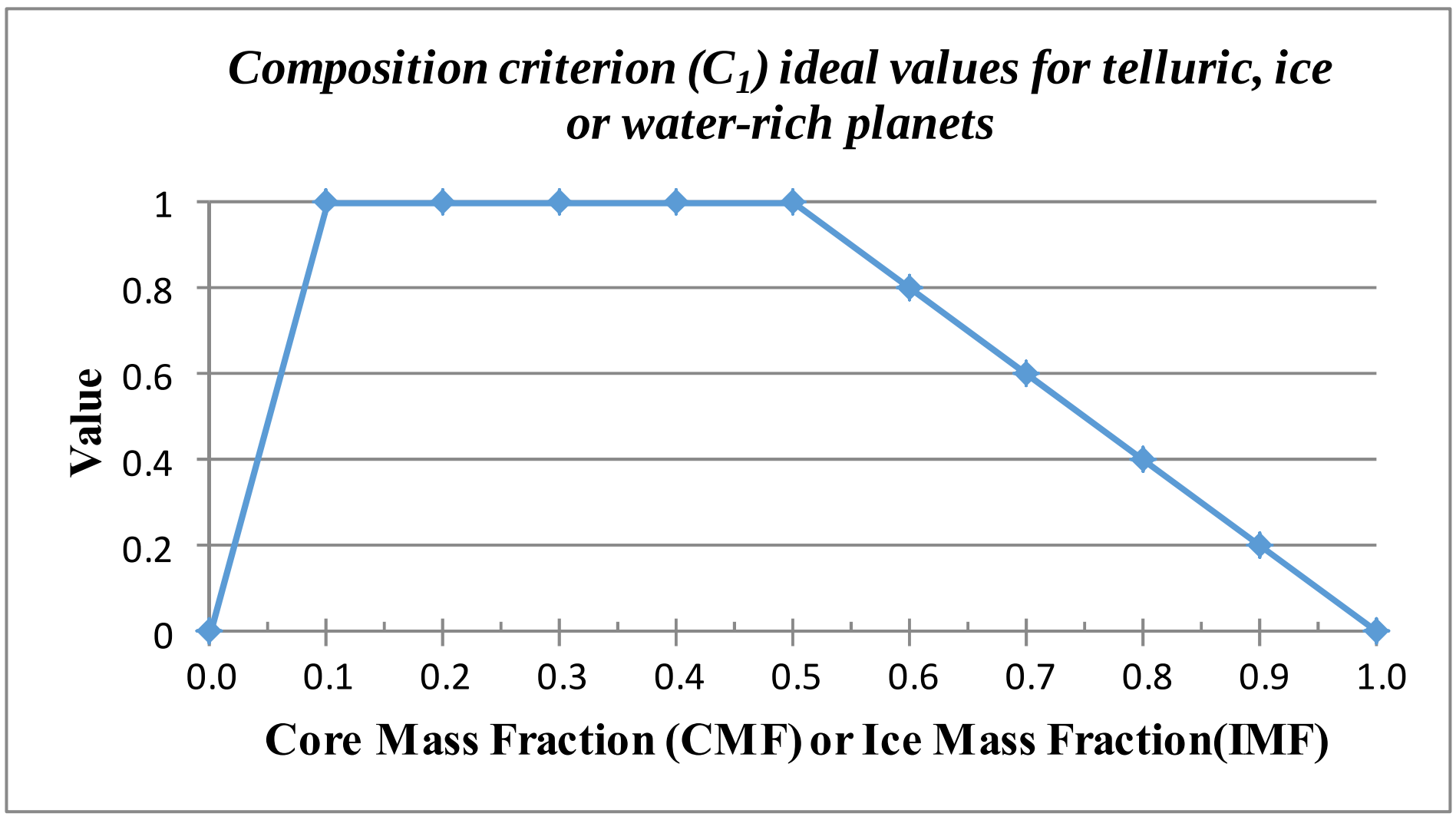

**Fig. 4:** Graphical Representation of the ideal values for Composition criterion ($C_1$) for telluric, ice or water-rich exoplanets (**Note**: Slopes and the start and end points of the ascending and descending lines respectively are arbitrary)

The ideal values of the escape velocity ($v_e$) in the atmosphere criterion ($C_2$) are based on the justification provided by Rodríguez-Mozos & Moya (2017). Since a very low gravity does not retain the atmosphere, and a very large value make chemical reactions and the formation of the structures necessary for life as we know very difficult, we adopt an aptitude conservative for this criterion. Therefore and according to Rodríguez-Mozos & Moya (2017), we define such ideal range with the aim of obtaining a 70% probability for an exoplanet having a scape velocity compatible with the development of life. Due to that, the ideal values of the escape velocity should be: $[C, D] = [0.8, 3]$.

With respect to the ideal values related to the energy that the planet receives from its star, we use the normalized effective stellar flux parameter (criterion $C_3$). That flux should be sufficient to keep water in a liquid state. For an Earth-like planet orbiting around the Sun, this range of fluxes varies between 1.11 and 0.36 (Kopparapu et al., 2013). However, the mass of the planet varies the limits of the habitable zone, so that more massive planets can keep water in a liquid state closer to its star and less massive planets must be further away (Kopparapu et al., 2014). For this reason, we have done a conservative estimation reducing the maximum ideal flow by a 5%, reaching the ideal maximum value of 1.05.

Furthermore, on geologically active planets, the carbonate-silicate cycle can stabilize the level of carbon dioxide in the atmosphere and cause terrestrial-type planets to keep water in a liquid state. The limit of the outer zone of HZ is marked by the condensation point of $CO_2$, disappearing the greenhouse effect that finally leads the planet to a snowball state (Selsis et al, 2007).

The effective stellar flux that we have estimated reflects on the one hand the upper part of the range indicated above, and on the other hand atmospheres rich in water vapor similar to the Earth, in contrast with lower effective fluxes than would entail a significant presence of greenhouse gases in their atmospheres. Therefore, $[C, D] = [0.7, 1.05]$ represents the estimated ideal values of the effective stellar flux related to the energy criterion ($C_3$).

The magnetic moment of The Earth has allowed the beginning and evolution of life. For this reason, we take as a reference the value of its magnetic moment. Because those planets having a magnetic moment equal to or greater than Earth enable to efficiently protect their potential life, we can estimate the range $[C, D] = [1, 10]$ as the ideal values of the normalized magnetic moment (criterion $C_4$). Values close to zero imply low-mass and/or slow rotating planets which would have weak multipolar magnetic moments, not appropriate for protecting life (Rodríguez-Mozos & Moya, 2017).

Finally, the ideal reference values for the qualitative criteria $C_5$, $C_6$ and $C_7$ (Tidal locking zones, substrate and habitable zones and its boundaries respectively), have been designed in such a way that they enable to cover completely the linguistic labels best valued for these criteria, that is, unlocked and telluric worlds, and located within the green zone. So, $[\tilde{C}, \tilde{D}] = [(6.5, 7, 7.5), (8.5, 9.5, 10)]$ defines the ideal range of the $C_5$ (Tidal locking zones), $C_6$ (substrate) and $C_7$ (habitable zones and its boundaries) criteria.

### 3.2.2. Application of RIM and Fuzzy RIM methodologies

The final stage of this study is to obtain the normalized decision matrix through expressions (2) and (3) for quantitative criteria and (4), (5) and (6) for qualitative criteria (see Methodology Section). Then, the weighted normalized matrix is obtained by multiplying the values of each element of the normalized decision matrix by the weight of each criterion via expression (7).

The RIM methodology allows to assign a weight or coefficient of importance for each criterion. Such weights can be obtained by applying other MCDM methodologies such as the fuzzy AHP approach through the extraction of knowledge from a group of experts. However, with the aim of not adding more subjectivity to the proposed case study, we start from the basis that all criteria have an influence in the assessment process of equal importance from the point of view of exoplanetary habitability. Nevertheless, two exceptions are worth noting:

The first one involves criterion $C_3$ (Normalized effective stellar flux) and criterion $C_7$ (Habitable zones and its boundaries). Although such criteria integrate additional information and assumptions, both are strongly correlated. To ensure that any possible dependency relationships between criteria having less influence than the impact of the individual criteria, we have weighted these two criteria with a factor one half each.

The second exception is related to the Tidal locking criteria which involves criterion $C_4$ (Magnetic moment) and criterion $C_5$ (Tidal locking zones). Similarly to that mentioned for the case of criteria $C_3$ and $C_7$, the strong connection between criteria $C_4$ and $C_5$ implies that their weights must be distributed equitably. Therefore, the initial weight of each one of these criteria ($C_4$ and $C_5$) must be multiplied by 0.5.

Once obtained the weighted normalized decision matrix, the rest of the stages of the RIM algorithm (from stage 5 to stage 7) can be addressed, concluding with a ranking of exoplanets based on the ideal reference alternative i.e, in terms of their similarity to The Earth.

### 3.3. Results

As a result of the evaluation process of exoplanets based on their habitability and similarity to the Earth, 1798 exoplanets have been analyzed by means of the RIM approach and its Fuzzy version (Fuzzy RIM); this provides a relative index ($R_i$) for each one of the alternatives, obtaining a ranking of exoplanets from the perspective of the exoplanetary habitability.

Since it is not possible to represent the decision matrix fully in a clear and coherent manner, then, ranking RIM and the decision matrix of the 29 top-ranked exoplanets are shown (Table 8).

| *Exoplanet* | *($R_i$)* | *Ranking* | *$C_1$* | *$C_2$* | *$C_3$* | *$C_4$* | *$C_5$* | *$C_6$* | *$C_7$* |
|---|---|---|---|---|---|---|---|---|---|
| Kepler-442b | 0.902 | 1 | 0.21 | 1.35 | 0.60 | 1.93 | Potent. Locked | Telluric | Green |
| Kepler-443b | 0.822 | 2 | 0.40 | 1.62 | 0.83 | 4.38 | Potent. Locked | Ice | Green |
| Kepler-062f | 0.802 | 3 | 0.09 | 1.37 | 0.50 | 1.88 | Unlocked | Ice | Green |
| Kepler-174d | 0.774 | 4 | 0.58 | 1.57 | 0.43 | 3.85 | Unlocked | Ice | Green |
| Kepler-062e | 0.769 | 5 | 0.28 | 1.43 | 1.41 | 1.55 | Potent. Locked | Ice | Inner edge |
| TRAPPIST-1e | 0.751 | 6 | 0.32 | 0.92 | 0.69 | 0.11 | Tidally Locked | Telluric | Green |
| LHS_1140b | 0.746 | 7 | 0.36 | 2.00 | 0.48 | 0.26 | Tidally Locked | Telluric | Green |
| Kepler-1649c | 0.746 | 8 | 0.18 | 1.04 | 0.78 | 0.06 | Tidally Locked | Telluric | Green |
| Kepler-186f | 0.745 | 9 | 0.21 | 1.18 | 0.35 | 0.10 | Potent. Locked | Telluric | Green |
| TOI-700d | 0.745 | 10 | 0.20 | 1.25 | 0.87 | 0.05 | Tidally Locked | Telluric | Green |
| Kepler-283c | 0.741 | 11 | 0.42 | 1.48 | 0.89 | 0.07 | Potent. Locked | Ice | Green |
| Kepler-296f | 0.737 | 12 | 0.38 | 1.48 | 0.61 | 0.09 | Potent. Locked | Ice | Green |
| K2-072e | 0.737 | 13 | 0.31 | 0.76 | 0.79 | 0.02 | Tidally Locked | Telluric | Green |
| Kepler-069c | 0.732 | 14 | 0.39 | 1.43 | 2.05 | 2.37 | Unlocked | Ice | Inner edge |
| Kepler-439b | 0.720 | 15 | 0.41 | 1.65 | 1.81 | 3.41 | Potent. Locked | Ice | Inner edge |
| Kepler-538 | 0.720 | 16 | 0.25 | 2.21 | 2.95 | 4.39 | Potent. Locked | Ice | Inner edge |
| Kepler-440b | 0.719 | 17 | 0.50 | 1.48 | 1.36 | 0.18 | Potent. Locked | Ice | Inner edge |
| Kepler-441b | 0.714 | 18 | 0.26 | 1.43 | 0.20 | 2.20 | Unlocked | Ice | Outer edge |
| Kepler-220e | 0.711 | 19 | 0.14 | 1.33 | 3.61 | 0.06 | Potent. Locked | Telluric | Inner edge |
| K2-009b | 0.702 | 20 | 0.27 | 1.42 | 1.15 | 0.24 | Tidally Locked | Ice | Inner edge |
| Kepler-061 | 0.696 | 21 | 0.62 | 1.57 | 1.25 | 0.18 | Potent. Locked | Ice | Inner edge |
| Kepler-367c | 0.678 | 22 | 0.26 | 1.23 | 3.33 | 0.03 | Tidally Locked | Telluric | Inner edge |
| Kepler-155c | 0.677 | 23 | 0.51 | 1.59 | 2.44 | 0.21 | Potent. Locked | Ice | Inner edge |
| Kepler-437b | 0.674 | 24 | 0.47 | 1.55 | 2.07 | 0.15 | Potent. Locked | Ice | Inner edge |
| Kepler-296e | 0.673 | 25 | 0.22 | 1.39 | 1.40 | 0.11 | Tidally Locked | Ice | Inner edge |
| K2-003d | 0.652 | 26 | 0.26 | 1.43 | 1.60 | 0.11 | Tidally Locked | Ice | Inner edge |
| TRAPPIST-1g | 0.644 | 27 | 0.09 | 1.01 | 0.27 | 0.11 | Tidally Locked | Ice | Green |
| TRAPPIST-1f | 0.588 | 28 | 0.02 | 0.94 | 0.40 | 0.11 | Tidally Locked | Telluric | Green |
| TRAPPIST-1h | 0.578 | 29 | 0.09 | 0.65 | 0.15 | 0.02 | Tidally Locked | Ice | Outer edge |

**Table 8:** Relative Index ($R_i$), Ranking RIM and Decision matrix of the twenty nine Top-ranked alternatives

This ranking can be compared with that provided by the SEPHI index. In Table 9 we show this comparison.

| *Exoplanet* | *($R_i$)* | *FRIM* | *SEPHI* |
|---|---|---|---|
| Kepler-442b | 0.90232 | 1 | 1 |
| Kepler-443b | 0.82157 | 2 | 40 |
| Kepler-062f | 0.80208 | 3 | 2 |
| Kepler-174d | 0.77420 | 4 | 37 |
| Kepler-062e | 0.76865 | 5 | 4 |
| TRAPPIST-1e | 0.75143 | 6 | 11 |
| LHS_1140b | 0.74643 | 7 | 5 |
| Kepler-1649c | 0.74625 | 8 | 16 |
| Kepler-186f | 0.74518 | 9 | 7 |
| TOI-700d | 0.74514 | 10 | 15 |
| Kepler-283c | 0.74076 | 11 | 31 |
| Kepler-296f | 0.73743 | 12 | 19 |
| K2-072e | 0.73712 | 13 | 26 |
| Kepler-069c | 0.73226 | 14 | 6 |
| Kepler-439b | 0.72006 | 15 | 8 |
| Kepler-538 | 0.72006 | 16 | 14 |
| Kepler-440b | 0.71910 | 17 | 28 |
| Kepler-441b | 0.71449 | 18 | 3 |
| Kepler-220e | 0.71058 | 19 | 46 |
| K2-009b | 0.70224 | 20 | 10 |
| Kepler-061 | 0.69642 | 21 | 62 |
| Kepler-367c | 0.67815 | 22 | 45 |
| Kepler-155c | 0.67702 | 23 | 135 |
| Kepler-437b | 0.67413 | 24 | 80 |
| Kepler-296e | 0.67258 | 25 | 23 |
| K2-003d | 0.65196 | 26 | 38 |
| TRAPPIST-1g | 0.64435 | 27 | 13 |
| TRAPPIST-1f | 0.58800 | 28 | 12 |
| TRAPPIST-1h | 0.57831 | 29 | 32 |

**Table 9:** Comparison of the twenty nine Top-ranked alternatives based on FRIM and SEPHI

We have not compared with other indexes such as ESI because this index doesn't take into account the planetary magnetic field, and it prioritizes exoplanets orbiting VLM stars, independently whether it is tidally locked or not. On the other hand, FRIM and SEPHI include this physics, and every tidally locked planet is penalized and it shrinks down in the ranking. Therefore, the results coming from these indexes are not comparable.

Looking at the comparison of Table 9, we find that the first planet in both rankings is the same: Kepler-442 b. Therefore, we can ensure that it is the perfect candidate for searching for biomarkers in its atmosphere, if technically possible. Bellow this planet we find similar rakings for Kepler-62 f, Kepler-62 e, Kepler-186 f, or LHS 1140 b, for example. All in the top ten of both indexes. On the other hand, we also find remarkable differences, as it is the case of Kepler-443 b and K2-72 e. Most of them related to the compensatory nature of FRIM i.e., a not very good value in one criterion can be compensated by better values in other criteria. Furthermore, SEPHI values dry planets directly with a probability of 1 whereas in FRIM dry planets are valued through CMF and those rich in water with IMF.

Another remarkable case is TRAPPIST-1 e. Both indexes rank this exoplanet in positions 6 and 11, making it a very good candidate for further analysis. In any case, we must take into account that SEPHI is not a ranking algorithm. It provides an statistical likelihood of habitability. Therefore, its eleventh position in SEPHI ranking is not that meaningful. Its SEPHI index is 0.4, that is, it has a probability of a 40% of being as habitable as the Earth. Taking into account its proximity to the earth and its VLM host star, its prioritization for searching for biomarkers seems reasonable, since this goal it is technically feasible in short term and its habitability likelihood and FRIM ranking are significant.

## 4. Conclusions

With the aim of offering tools for the prioritization of known exoplanets for searching for biomarkers, we have presented a fuzzy MCDM approach to solve this goal. Fuzzy RIM is an algorithm based on MCDM methodologies that not only allows multiple exoplanets to be evaluated based on an ideal reference alternative, as is the case of our planet, but also allows for taking into account several criteria simultaneously which can present different nature (qualitative and quantitative criteria). For the implementation of this algorithm, we have defined seven criteria covering all the aspects making an exoplanet a potential habitable candidate. These criteria are the exoplanet composition, the presence of an atmosphere, energy input, two about tidal locking, planet type from its mean density, and the presence of liquid water on its surface (the so called Habitable Zone). All of these criteria are those with a larger impact in potential habitability that can be estimated using physical quantities provided by most of the past, current and future planet discovery missions.

Training this algorithm with these criteria and taking the Earth as reference or success case we are able to rank any observed exoplanet. We have used the TEPCat data base as reference for testing the FRIM. Comparing the results found with the habitability index SEPHI, we found very interesting results.

Kepler-442 b, Kepler-062 e/f, LHS_1140b and Kepler-186 f are the best exoplanets for searching for biomarkers, regardless its technical difficulty, since they appears consistently in the top ten ranking of both indexes. But if we add the technical feasibility, the best candidate is TRAPPIST-1 e, with a consistent ranking with both indexes and an habitability likelihood, coming from SEPHI, of a 40%.

## Acknowledgement

This research has been partially funded by the research project (PGC2018-097374-B-I00), funded by FEDER/Ministerio de Ciencia e Innovación – Agencia Estatal de Investigación. J.M.S.L. also acknowledges the support of Grants TIN2017-86647-P and 19882-GERM- 15 from the Spanish Ministry of Economy and Competitiveness (MINECO) and Fundación Séneca (Región de Murcia), respectively. AM acknowledges funding support from Spanish public funds (including FEDER fonds) for research under project PID2019-107061GB-C65. The authors sincerely acknowledge the referees for their fruitful and constructive comments.

## References

1. Baglin, A., Auvergne, M., Barge, P., et al., 2006. *The CoRoT Mission, PreLaunch Status, Stellar Seismology and Planet Finding*, ed. M. Fridlund, A. Baglin, J. Lochard, & L. Conroy, ESA Publications Division, Nordwijk, Netherlands, ESA SP-1306, 33.
2. Barnes, R., Meadows, V.S., Evans, N., 2015. Comparative Habitability of Transiting Exoplanets. *The Astrophysical Journal* 814(2):1-11.
3. Basak, S., Saha, S., Mathur, A., Bora, K., Makhija, S., Safonova, M., Agrawal, S., 2020. CEESA meets machine learning: A Constant Elasticity Earth Similarity Approach to habitability and classification of exoplanets. *Astronomy and Computing* 30, 100335.
4. Bin, J., Tian, F., Liu, L., 2018. New inner boundaries of the habitable zones around M dwarfs. *Earth and Planetary Science Letters* 492: 121–129.

5. Bora, K., Saha, S., Agrawal, S., Safonova, M., Routh, S., Narasimhamurthy, A., 2016. CD-HPF: New habitability score via data analytic modeling. *Astronomy and Computing* 17: 129–143.

6. Borucki, W.J., Koch, D.G., Brown, T.M., et al., 2010. Kepler Planet-Detection Mission: Introduction and First Results. *Science* 327: 977-980.

7. Brans, J.P., Mareschal, B., Vincke, Ph., 1984. *PROMETHEE: A new family of outranking methods in multicriteria analysis*. In J. P. Brans, editor, North-Holland, Amsterdam, Operational Research, pp 477–490.

8. Cables, E., Lamata, M.T., Verdegay, J.L., 2016. RIM-reference ideal method in multicriteria decision making. *Information Sciences* 337-338: 1-10.

9. Cables, E., Lamata, M.T., Verdegay, J.L., 2018. Soft Computing Applications for Group Decision-making and Consensus Modeling: FRIM—*Fuzzy Reference Ideal Method in Multicriteria Decision Making*. Mikael Collan Janusz Kacprzyk Editors- Springer International Publishing AG, pp 305-317.

10. Chen, J., Kipping, D., 2017. Probabilistic forecasting of the Masses and Radii of other worlds. *The Astrophysical Journal* 834 (17):1-13.

11. Dressing, C.D., Charbonneau, D., Dumusque, X., Gettel, S., Pepe, F., Cameron, A.C., Latham, D.W. et al., 2015. The mass of Kepler-93b and the composition of terrestrial planets. *The Astrophysical Journal* 800 (135):1-7.

12. Dubois, D., Prade, H., 1980. *Fuzzy Sets and Systems: Theory and Applications*, New York, NY, USA: Academic Press, 1980.

13. Fortier, A. et al., 2014. CHEOPS: a space telescope for ultra-high precision photometry of exoplanet transits. Space Telescopes and Instrumentation 2014: Optical, Infrared, and Millimeter Wave 9143.

14. Fortney, J.J., Marley, M.S., Barnes, J.W., 2007. Planetary Radii across Five Orders of Magnitude in Mass and Stellar Insolation: Application to Transits. *The Astrophysical Journal* 659 (2): 1661–1672.

15. Frederick, J.E., Lubin, D., 1988. The Budget of Biologically Active Ultraviolet Radiation in the Earth-Atmosphere System. *Journal of Geophysical Research* 93 (4): 3825-3832.

16. Gaidos, E., Deschenes, B., Dundon, L., Fagan, K., Mcnaughton, C., Menviel-Hessler, L., Moskovitz, N., Workman, M., 2005. Beyond the Principle of Plentitude: A Review of Terrestrial Planet Habitability. *Astrobiology* 5(2): 100-126.

17. García-Cascales, M.S., Lamata, M.T., Sánchez-Lozano, J.M., 2012. Evaluation of photovoltaic cells in a multi-criteria decision making process. *Annals of Operations Research* 199(1): 373–391.

18. Gebauer, S., 2003. *Dissertation: Evolution of Earth-like extrasolar planetary atmospheres*. Mathematik und Naturwissenschaften der Technischen Universität Berlin.

19. Gebauer, S., Grenfell, J.L., Stock, J.W., Lehmann, R., Godolt, M., von Paris, P., Rauer, H., 2017. Evolution of Earth-like Extrasolar Planetary Atmospheres: Assessing the Atmospheres and Biospheres of Early Earth Analog Planets with a Coupled Atmosphere Biogeochemical Model. *Astrobiology* 16: 27–54.

20. Grießmeier, J.-M., Stadelmann, A., Motschmann, U., Belisheva, N.K., Lammer, H., Biernat, H.K., 2005. Cosmic Ray Impact on Extrasolar Earth-Like Planets in Close-in Habitable Zones. *Astrobiology* 5: 587-603.

21. Grießmeier, J.-M., Stadelmann, A., Grenfell, J.L., Lammer, H., Motschmann, U., 2009. On the protection of extrasolar Earth-like planets around K/M stars against galactic cosmic rays. *Icarus* 199: 526-525.

22. Hart, M.H., 1979. Habitable zones about main sequence stars. *Icarus* 37: 351–357.

23. Heath, M.J., Doyle, L.R., Joshi, M.M., Haberbe, R.M., 1999. Habitability of planets around red dwarf stars. Origins Life Evolut. *Biosphere* 29: 405–424.

24. Heller, R., Armstrong, J., 2014. Superhabitable worlds. *Astrobiology* 14: 50–66.

25. Hu, Y., Yang, J., 2014. Role of ocean heat transport in climates of tidally locked exoplanets around M dwarf stars. *Proceedings of the National Academy of Sciences* 111 (2): 629–634.

26. Huang, S.S., 1960. Life outside the Solar System. *American Scientist* 202(4): 55–63.

27. Hwang, C.L., Yoon, K., 1981. *Multiple Attribute Decision Methods and Applications.* Springer, Berlin Heidelberg.

28. Irwin, L.N., Schulze-Makuch, D., 2001. Assessing the Plausibility of Life on Other Worlds. *Astrobiology* 1(2): 143-160.

29. Irwin, L.N., Schulze-Makuch, D., 2011. *Cosmic Biology: How Life Could Evolve on Other World*. Springer-Praxis, New York.

30. Irwin, L.N., Méndez, A., Fairén A.G., Schulze-Makuch, D., 2014. Assessing the Possibility of Biological Complexity on Other Worlds, with an Estimate of the Occurrence of Complex Life in the Milky Way Galaxy. *Challenges* 5: 159-174.

31. Joshi, M.M., Haberle, R.M., Reynolds, R.T., 1997. Simulations of the Atmospheres of Synchronously Rotating Terrestrial Planets Orbiting M Dwarfs:

Conditions for Atmospheric Collapse and the Implications for Habitability. *Icarus* 129: 450–465

32. Kaltenegger, L., Sasselov, D., 2011. Exploring the habitable zone for Kepler planetary candidates. *The Astrophysical Journal Letters* 736:1-6.

33. Kashyap, J.M., Gudennavar, S.B., 2017. *Dissertation: Earth Similarity Index and Habitability Studies of Exoplanets*. Department of Physics and Electronics, Christ University, Bengaluru, India.

34. Kashyap, J.M., Gudennavar, S.B., Doshi, U., Safonova. M., 2017. Similarity indexing of exoplanets in search for potential habitability: application to Mars-like worlds. *Astrophysics and Space Science* 362(146):1-13.

35. Kasting, J.F., 1988. Runaway and moist greenhouse atmospheres and the evolution of Earth and Venus. *Icarus* 74:472–494.

36. Kasting, J.F., 1993. Earth's early atmosphere. *Science* 259: 920–926.

37. Kasting, J.F. and Catling, D., 2003. Evolution of a habitable planet. *Annu. Rev. Astron. Astrophys*. 41:429–463.

38. Kasting, J.F., Siefert, J.L., 2002. Life and the Evolution of Earth's Atmosphere. *Science* 296 (5570): 1066-1068.

39. Kasting, J.F., Whitmire, D.P., Reynolds, R.T., 1993. Habitable zones around Main Sequence Stars. *Icarus* 101: 108-128.

40. Klir, G.J., Yuan, B., 1995. *Fuzzy sets and fuzzy logic: Theory and applications*. Upper Saddle River, NJ , USA: Prentice Hall. Upper Saddle River, NJ (1995).

41. Kopparapu, R.K., Ramirez, R., Kasting, J.F., Eymet, V., Robinson, T.D., Mahadevan, S., Terrien, R.C., Domagal-Goldman, S., Meadows, V., Deshpande, R., 2013. Habitable zones around Main-Sequence Stars: New estimates. *The Astrophysical Journal* 765(131): 1-16

42. Kopparapu, R.K., Ramirez, R.M., SchottelKotte, J., Kasting, J.F., Domagal-Goldman, S., Eymet, V., 2014. Habitable zones around Main-Sequence Stars: Dependence on planetary mass. *The Astrophysical Journal Letters* 787 (29): 1-6.

43. Laarhoven, P.J.M. Pedrycz, W., 1983. A Fuzzy Extension of Saaty's Priority Theory. *Fuzzy Sets and Systems* 11: 229-241.

44. Lammer, H., Bredehöft, J.H., Coustenis, A., Khodachenko, M.L., Kaltenegger, L., Grasset, O., Prieur, D., Raulin, F., Ehrenfreund, P., Yamauchi, M., Wahlund, J.-E., Grießmeier, J.-M., Stangl, G., Cockell, C.S., Kulikov, Y.N., Grenfell, J.L., Raue, H., 2009. What makes a planet habitable?. *Astron Astrophys Rev* 17:181–249.

45. Lammer, H., Selsis, F., Chassefie`re, E., Breuer, D., et al., 2010. Geophysical and Atmospheric Evolution of Habitable Planets. *Astrobiology* 10 (1): 45-68.

46. Leger, A., Pirre, M., Marceau, F.J., 1993. Search for primitive life on a distant planet: Relevance of O2 and O3 detections. *Astronomy & Astrophysics* 277: 309–313.

47. Madhusudhan, N., Agúndez, M., Moses, J.I., Hu, Y., 2016. Exoplanetary Atmospheres—Chemistry, Formation Conditions, and Habitability. *Space Sci Rev* 205:285–348.

48. Mardani, A. Jusoh, A., Nor, K.M.D., Khalifah, Z., Zakwan, N., Valipour, A., 2015. Multiple criteria decision-making techniques and their applications – a review of the literature from 2000 to 2014. *Economic Research-Ekonomska Istraživanja* 28(1): 516–571.

49. Mayor, M., Queloz, D., 1995. A Jupiter-mass companion to a solar-type star. *Nature* 378 (6555): 355–359.

50. Opricovic, S., 1998. *Multi-Criteria Optimization of Civil Engineering Systems*, Faculty of Civil Engineering, Belgrade.

51. Rauer, H. et al., 2014. The PLATO 2.0 mission. *Experimental Astronomy* 38, 249–330.

52. Rekola, R.T.F., 2009. Life and habitable zones in the Universe. *Planetary and Space Science* 57: 430–433.

53. Ricker, G.R., et al., 2014. Transiting Exoplanet Survey Satellite (TESS). *In Space Telescopes and Instrumentation 2014: Optical, Infrared, and Millimeter Wave*, vol. 9143 of SPIE Proceedings, 914320. 1406.0151.

54. Rodríguez-Mozos, J.M., Moya, A., 2017. Statistical-likelihood Exo-Planetary Habitability Index (SEPHI). *Monthly Notices of the Royal Astronomical Society* 471 (4):4628–4636.

55. Rodríguez-Mozos, J.M., Moya, A., 2019. Erosion of an exoplanetary atmosphere caused by stellar winds, *Astronomy & Astrophysics* 630 (52): 1-12.

56. Roy, B., 1968. Classement et choix en présence de points de vue multiples (la méthode ELECTRE). *La Revue d'Informatique et de Recherche Opérationelle (RIRO)* 8: 57–75.

57. Saaty, T.L., 1980. *The Analytic Hierarchy Process*. New York, USA: McGraw Hill International.

58. Saaty, T.L., 1996. *Decision Making With Dependence and Feedback: The Analytic Network Process*. Pittsburgh, PA, USA: RWS Publisher.

59. Sánchez-Lozano, J.M., Fernández-Martínez, M., 2016. Near-Earth object hazardous impact: A Multi-Criteria Decision Making approach. *Scientific Reports* 6, 37055:1-10.

60. Sánchez-Lozano, J.M., Naranjo-Rodríguez, O., 2020. Application of Fuzzy Reference Ideal Method (FRIM) to the military advanced training aircraft selection. *Applied Soft Computing Journal* 88, 106061.

61. Sánchez-Lozano, J.M., Fernández-Martínez, M., Lamata, M.T., 2019. Near-Earth Asteroid impact dates: A Reference Ideal Method (RIM) approach. *Engineering Applications of Artificial Intelligence* 81: 157-168.

62. Schulze-Makuch, D., Méndez, A., Fairén, A.G., von Paris, P., Turse, C., Boyer, G., Davila, A.F., de Sousa António, M.R., Catling, D., Irwin, L.N., 2011. A Two-Tiered Approach to Assessing the Habitability of Exoplanets. *Astrobiology* 11 (10): 1041-1052.

63. Seager, S., Kuchner, M., Hier-Majumder, C.A., Militzer, B., 2007. Mass-radius relationships for solid exoplanets. *The Astrophysical Journal* 669:1279-1297.

64. Selsis, F., Kasting, J.F., Levrard, B., Paillet, J., Ribas, I., Delfosse, X., 2007. Habitable planets around the star Gliese 581?. *Astronomy & Astroph*ysics 476: 1373–1387.

65. Simon, H., 1960. *The New Science of Management Decision*. New York: Harper and Row.

66. Southworth, J., 2020. *TEPCat: Physical properties of transiting planets*. Keele University, UK. https://www.astro.keele.ac.uk/jkt/tepcat/allplanets-noerr.html [accessed 03.03.20]

67. Suissa, G., Chen, J., Kipping, D., 2018. A HARDCORE model for constraining an exoplanet's core size. *Monthly Notices of the Royal Astronomical Society* 476 (2): 2613–2620.

68. Sundquist, E.T., 1993. The global carbon dioxide budget. *Science* 259: 934-941.

69. Tavana, M., 2003. CROSS: A multiattributes group-decision-making model for evaluating and prioritizing advanced-technology projects at NASA. *Interfaces* 33(3): 40–56.

70. Tavana, M., 2004. A subjective assessment of alternative mission architectures for the human exploration of Mars at NASA Using multiattributes decision making. *Computers and Operations Research* 31(7): 1147–1164.

71. Tavana, M., Hatami-Marbini, A., 2011. A group AHP-TOPSIS framework for human spaceflight mission planning at NASA. *Expert Systems with Applications* 38: 13588–13603.

72. Triantaphyllou, E., 2000. *Multi-Criteria Decision Making Methods: A comparative study*. Dordrecht, The Netherlands: Kluwer Academic Publishers.

73. Truitt, A., Young, P.A., Spacek, A., Probst, L., Dietrich, J., 2015. A catalog of stellar evolution profiles and the effects of variable composition on habitable systems. *The Astrophysical Journal* 804 (2): 1-16.

74. Valencia, D., O'Connell, R.J., Sasselov D., 2006. Internal structure of massive terrestrial planets. *Icarus* 181(2): 545-554.

75. Wallace, A.R., 1903. *Man's Place in the Universe*, McClure, Phillips and Co., New York.

76. Wolstencroft, R.D., Raven, J.A., 2002. Photosynthesis: likelihood of occurrence and possibility of detection on Earth-like planets. *Icarus* 157: 535–548.

77. Wolszczan, A., Frail, D., 1992. A planetary system around the millisecond pulsar PSR1257 + 12. *Nature* 355 (6356): 145–147.

78. Zadeh, L.A., 1965. Fuzzy sets. *Information and Control* 8: 338–353.

79. Zadeh, L.A., 1975. The concept of linguistic variable and its application to approximate reasoning. *Information Sciences* 8: 199-249.

80. Zeng, L., Sasselov. D., 2013. A Detailed Model Grid for Solid Planets from 0.1 through 100 Earth Masses. *Publications of the Astronomical Society of the Pacific* 125 (925):1-13.

81. Zuluaga, J.I., Bustamante, S., 2018. Magnetic properties of Proxima Centauri b analogues. *Planetary and Space Science* 152: 55-67.